\documentclass[traditabstract]{aa} 

\usepackage{natbib}
\usepackage{amsmath}
\bibpunct{(}{)}{;}{a}{}{,}

\usepackage{graphicx}
\usepackage{txfonts}
\usepackage{url}

\newcommand{\feh}{\mbox{[Fe/H]}}
\newcommand{\teff}{\mbox{$T_{\rm eff}$}}
\newcommand{\logg}{\mbox{$\log g_*$}}
\newcommand{\vsini}{\mbox{$v \sin I$}}

\newcommand{\mactrb}{\mbox{$v_{\rm mac}$}}

\newcommand{\kms}{\mbox{km\,s$^{-1}$}}
\newcommand{\ms}{\mbox{m\,s$^{-1}$}}

\newcommand{\mplanet}{\mbox{$M_{\rm pl}$}}
\newcommand{\rplanet}{\mbox{$R_{\rm pl}$}}
\newcommand{\densplanet}{\mbox{$\rho_{\rm pl}$}}
\newcommand{\loggplanet}{\mbox{$\log g_{\rm pl}$}}
\newcommand{\mjup}{\mbox{$M_{\rm Jup}$}}
\newcommand{\rjup}{\mbox{$R_{\rm Jup}$}}
\newcommand{\densjup}{\mbox{$\rho_{\rm Jup}$}}
\newcommand{\mstar}{\mbox{$M_*$}}
\newcommand{\rstar}{\mbox{$R_*$}}
\newcommand{\densstar}{\mbox{$\rho_*$}}

\newcommand{\msol}{\mbox{$M_\odot$}}
\newcommand{\rsol}{\mbox{$R_\odot$}}
\newcommand{\denssol}{\mbox{$\rho_\odot$}}

\def\secos{$\sqrt{e} \cos \omega$}
\def\sesin{$\sqrt{e} \sin \omega$}

\def\svsicos{$\sqrt{v \sin I} \cos \lambda$}
\def\svsisin{$\sqrt{v \sin I} \sin \lambda$}
\def\teql{$T_{\rm eql}$}

\mathchardef\mhyphen="2D

\def\rhk{$\log R'_{\rm HK}$}
\def\smw{$S_{\rm MW}$}

\begin{document}

\title{Spin-orbit measurements and refined parameters for the exoplanet systems 
WASP-22 and WASP-26
\thanks{
Based on observations made with the HARPS spectrograph on the 3.6-m ESO 
telescope (proposal 085.C-0393), the 0.6-m Belgian TRAPPIST telescope, and the 
CORALIE spectrograph and the Euler camera on the 1.2-m Euler Swiss telescope, 
all at the ESO La Silla Observatory, Chile.}
$^{, }$
\thanks{The photometric time-series and radial velocity data used in this 
work are only available in electronic form at the CDS via anonymous ftp to 
cdsarc.u-strasbg.fr (130.79.128.5) or via 
http://cdsweb.u-strasbg.fr/cgi-bin/qcat?J/A+A/}}

\titlerunning{Spin-orbit measurements of WASP-22b and WASP-26b}

\author{D.~R.~Anderson
        \inst{1}
	\and
A.~Collier~Cameron
        \inst{2}
	\and 
M.~Gillon
        \inst{3}
	\and 
C.~Hellier
        \inst{1}
	\and 
E.~Jehin
        \inst{3}
	\and 
M.~Lendl
        \inst{4}
	\and 
D.~Queloz
        \inst{4}
	\and 
B.~Smalley
        \inst{1}
	\and 
A.~H.~M.~J.~Triaud
        \inst{4}
	\and
M.~Vanhuysse
        \inst{5}
}

\institute{Astrophysics Group, Keele University, Staffordshire ST5 5BG, UK\\
           \email{dra@astro.keele.ac.uk}
           \and
           SUPA, School of Physics and Astronomy, University of St. Andrews, 
           North Haugh, Fife KY16 9SS, UK
	   \and
	   Institut d'Astrophysique et de G\'eophysique,  Universit\'e de 
	   Li\`ege,  All\'ee du 6 Ao\^ut, 17,  Bat.  B5C, Li\`ege 1, Belgium
	   \and
           Observatoire de Gen\`eve, Universit\'e de Gen\`eve, 51 Chemin 
           des Maillettes, 1290 Sauverny, Switzerland
	   \and
	   Oversky, 47 All\'ee des Palanques, 33127 Saint Jean d'Illac, France
}

\date{Received June 30, 2011; accepted September 1, 2011}
\authorrunning{D. R. Anderson et al.}

 
\abstract
{We report on spectroscopic and photometric observations through transits of the 
exoplanets WASP-22b and WASP-26b, intended to determine the systems' spin-orbit 
angles. 
We combine these data with existing data to refine the system parameters. 
We measure a sky-projected spin-orbit angle of $22 \pm 16^\circ$ for WASP-22b, 
showing the planet's orbit to be prograde and, perhaps, slightly misaligned. 
We do not detect the Rossiter-McLaughlin effect of WASP-26b due to its low 
amplitude and observation noise. 
We place 3-$\sigma$ upper limits on orbital eccentricity of 0.063 for WASP-22b 
and 0.050 for WASP-26b. 
After refining the drift in the systemic velocity of WASP-22 found by 
Maxted et al. (2010, AJ, 140, 2007), we find the third body in the system to 
have a minimum-mass of 5.3 $\pm$ 0.3 \mjup\ ($a_3$/5 AU)$^{2}$, 
where $a_3$ is the orbital distance of the third body.
}


\maketitle

%

\section{Introduction}
By taking spectra of a star whilst an exoplanet transits across it we can 
measure the sky-projected obliquity $\lambda$, where obliquity is the angle 
between the stellar rotation axis and the planetary orbital axis. 
As the planet obscures a portion of the rotating star it causes a distortion of  
the observed stellar line profile, which manifests as an anomalous 
radial-velocity (RV) signature known as the Rossiter-McLaughlin (RM) effect 
\citep{1893A&A....12..646H,1924ApJ....60...15R,1924ApJ....60...22M}. 
The shape of the RM effect is sensitive to the path a planet takes across the 
disc of a star relative to the stellar spin axis, thus by observing it we can 
determine $\lambda$. 
If we can determine the stellar inclination with respect to the sky plane $I$ 
then we can determine the system's true obliquity $\Psi$. 
For example, the light curve of a photospherically active star may exhibit 
rotational modulation, from which the true stellar rotation velocity can be 
determined. This can be compared with the spectroscopically-determined, 
sky-projected rotation velocity to get the stellar inclination 
\citep[e.g.][]{2007AJ....133.1828W, 2011ApJ...730L..31H}.

It is thought that the obliquity of a short-period, giant planet is indicative 
of the manner by which it arrived in its current orbit from farther out, where 
it presumably formed. 
As the angular momenta of a star and its planet-forming disc both derive from 
that of their parent molecular cloud, stellar spin and planetary orbital axes 
are expected to at least initially be aligned. 
Migration via tidal interaction with the gas disc is expected to preserve this 
initial spin-orbit alignment \citep{1996Natur.380..606L,2009ApJ...705.1575M}, 
but a significant fraction of the few dozen systems so far 
measured\footnote{Ren\'e Heller maintains a list of measurements and references 
at \url{http://www.aip.de/People/rheller/content/main_spinorbit.html}} are 
misaligned and some are even retrograde 
\citep[e.g. see][and the references therein]{2011A&A...527L..11H}. 
\citet{2010A&A...524A..25T} interpreted this as indicating that some or all 
close-in planets arrive in their orbits by a mixture of: planet-planet 
scattering and the Kozai mechanism, which can drive planets into eccentric, 
misaligned orbits; and tidal friction, which circularises and shortens orbits 
\citep[e.g.,][]{2007ApJ...669.1298F,2008ApJ...678..498N,2010ApJ...725.1995M,
2011Natur.473..187N}.
\citet{2011ApJ...729..138M} found the current spin-orbit distribution to be 
well described by migration via a combination of such alignment-preserving and 
misaligning mechanisms.

\citet{2010ApJ...718L.145W} and \citet{2010ApJ...719..602S} showed that 
the stars in misaligned systems tend to be hot (\teff\ $>$ 6250 K). 
\citet{2010ApJ...718L.145W} suggested that this indicates either that planetary 
formation and migration mechanisms depend on stellar mass, or that cooler stars, 
with their larger convective zones, are able to more effectively realign orbits 
via tidal dissipation.
The number of possible exceptions to this trend is increasing, such as 
HAT-P-9 with \teff\ = 6350 K and $\lambda = -16 \pm 8^\circ$, though rather than 
having been realigned, such systems may simply have never been very misaligned 
\citep{2011A&A...533A.113M}.

In addition to the scattering route, it has been suggested that misaligned 
systems can arise when stars are misaligned with their discs 
\citep{2011MNRAS.412.2790L} or when discs are misaligned with their stars 
\citep{2010MNRAS.401.1505B}.

In this paper we present observations of spectroscopic and photometric transits 
of WASP-22b \citep[][hereafter M10]{2010AJ....140.2007M} and WASP-26b 
\citep[][hereafter S10]{2010A&A...520A..56S}, from which we attempt to determine 
the obliquities and refine the parameters of the systems.

WASP-22b is a 0.59-\mjup\ planet that transits its solar metallicity 
(\feh\ = $0.05 \pm 0.08$) G0V, $V$=12.0 host star every 3.53 days. 
M10 found a linear trend in the RV measurements of WASP-22 
suggestive of either a second planet, a low-mass M-dwarf or a white dwarf 
companion. 

WASP-26b is a 1.03-\mjup\ planet that transits its solar metallicity 
(\feh\ = $-0.02 \pm 0.09$) G0V, $V$=11.3 host star every 2.76 days. 
S10 found WASP-26 to be a visual double with a K-type star, separated in the 
sky-plane by 3\,800 AU, and used it to infer a common age of $6 \pm 2$ Gyr.

\section{Observations}

\subsection{Spectroscopic transits and orbits}
\label{sec:obs:spec}

\begin{table*}
\centering
\caption{Stellar parameters}
\label{tab:sp}
\begin{tabular}{lcccccccc}
\hline
Star	& \teff		& \logg		& \feh				& \vsini	& \rhk			& \smw		 & $P_{\rm rot}$ & $v$		\\
	& (K)		& (cgs)		&				& (\kms)	& 			& 		 & (days)	 & (\kms)	\\
\hline
WASP-22	& $6000 \pm 100$& $4.5 \pm 0.2$	& ~~~\,$0.05 \pm 0.08$\tablefootmark{a}	& $4.5 \pm 0.4$	& $-4.90 \pm 0.06$	& $0.17 \pm 0.1$ & $14 \pm 3$	 & $4.3 \pm 0.8$\\
WASP-26	& $5950 \pm 100$& $4.3 \pm 0.2$	& $-0.02 \pm 0.09$		& $3.9 \pm 0.4$	& $-4.98 \pm 0.07$	& $0.16 \pm 0.1$ & $19 \pm 3$	 & $3.5 \pm 0.5$\\
\hline
\end{tabular}
\tablefoot{The values of \teff, \logg\ and \feh\ are from M10 for WASP-22 and 
from S10 for WASP-26. The other parameters are from this paper.}
\tablefoottext{a}{Due to a typographical error in M10, \feh\ was reported as 
$-0.05 \pm 0.08$.}
\end{table*}

A spectroscopic transit comprises a large number of RVs taken in quick 
succession, during which the target star may have a specific activity level that 
could bias the measured systemic velocity and hence the determination of 
$\lambda$. 
Typically, the stellar rotation period is much longer than the orbital 
period. So, to first order, stellar activity would manifest as a slow variation 
in the apparent RV of the system's centre of mass, which is negligibly small on 
the time-scale of a transit. 
To mitigate against the potential effects of activity, we obtained spectra of 
the two host stars outside of transit on the nights of the measured RM effects, 
and to refine the planets' masses and orbital eccentricities we obtained spectra 
across the full orbits.

We obtained spectra of WASP-22 and WASP-26 using the HARPS spectrograph on the 
ESO 3.6-m telescope at La Silla. 
Calibration spectra of the thorium-argon lamp were taken at the start of each 
night, thus avoiding contamination of the stellar spectra by the lamp. 
This is made possible by the stability of HARPS, which is at the 1 m s$^{-1}$ 
night$^{-1}$ level. 

On 2009 Nov 19, we obtained 20 spectra of WASP-22 through a transit with 
HARPS. 
The exposures were 15--20 min and the signal-to-noise ratio (SNR) per pixel 
at 550 nm was in the range 16--34, with 20 being typical. 
During the sequence, the airmass of the target decreased from 1.11 to 1.01 and 
then increased to 1.70. 
We sampled the full 3.53-day orbit of WASP-22b by obtaining a pair of spectra 
per night, well-spaced in time, on the night of the transit and on each of the 
preceeding three nights. 
With exposures of 30 min for these eight spectra, the SNR was 24--48. 

On 2010 Sep 12, we obtained 24 spectra of WASP-26 through a transit with 
HARPS. 
The exposures were 10--15 min and the SNR per pixel at 550 nm was in the 
range 10--32, with 22 being typical. 
During the sequence, the airmass of the target decreased from 1.62 to 1.03 and 
then increased to 1.08. 
As the focus was improved at BJD (UTC) = 2\,455\,451.641 the SNR and measurement 
precision improved, despite a decrease in the exposure time from 15 to 10 min. 
We sampled the full 2.76-day orbit of WASP-26b by obtaining six spectra over a 
period of five days around the transit. 
With exposures of 15--20 min, the SNR of these spectra was 26--40. 

After removing the instrumental blaze function \citep{2010A&A...524A..25T}, 
radial-velocity (RV) measurements were computed for each star by weighted 
cross-correlation with a numerical G2-spectral template 
\citep{1996A&AS..119..373B,2005Msngr.120...22P}. 
These are available online only. 
We ascribe the fact that the first five RVs in the transit sequence of WASP-26 
are lower precision than the rest to higher airmass and poorer focus.

We also incorporated in our analysis the 37 CORALIE RVs and 6 HARPS spectra of 
WASP-22 reported in M10 and the 16 CORALIE RVs of WASP-26 reported in S10. 
For consistency with the new measurements, we recalculated RVs from the existing 
HARPS spectra of WASP-22 after removal of the instrumental blaze function, which 
had not been done previously. 
We present these recomputed RVs along with the new RVs in the online tables. 
The new and old RVs are plotted in Figure~\ref{fig:w22-rv-phot}a for WASP-22 
and in Figure~\ref{fig:w26-rv-phot}a for WASP-26, folded on the ephemerides of 
Table~\ref{tab:mcmc}. 

We estimated the sky-projected stellar rotation velocity\footnote{We use $I$ for 
the angle between the stellar spin axis and the sky plane and $i$ for the angle 
between the planetary orbital axis and the sky-projection of the stellar spin 
axis.} \vsini\ from the HARPS spectra by fitting the profiles of several 
unblended Fe~{\sc i} lines. 
For this we determined an instrumental broadening of $0.060 \pm 0.005$ \AA\
from telluric lines around 6300 \AA\ and used the \citet{2010MNRAS.405.1907B} 
calibration to assume values for macroturbulence: 
\mactrb\ = $3.2 \pm 0.3$ \kms\ for WASP-22 and 
\mactrb\ = $3.0 \pm 0.3$ \kms\ for WASP-26.
We obtained \vsini\ values of $4.5 \pm 0.4$ \kms\ for WASP-22 and $3.9 \pm 0.4$ 
\kms\ for WASP-26.

We determined the typical activity indices (\rhk\ and \smw) of each star by 
measuring the weak emission in the cores of the Ca {\sc ii} H+K lines in the 
HARPS spectra with the highest SNR per pixel at 550 nm: 13 spectra with 
SNR$>$32 for WASP-22 and 11 spectra with SNR$>$27 for WASP-26 
\citep{1984ApJ...287..769N, 2000A&A...361..265S, 2009A&A...495..959B}. 
The values given in Table~\ref{tab:sp} are the averages and standard deviations 
of the values from individual spectra, determined using $B-V$ values of 0.57 for 
WASP-22 and 0.59 for WASP-26. 
For each star, we estimated the true stellar-rotation period, $P_{\rm rot}$, 
using the activity-rotation calibration of \citet{2008ApJ...687.1264M}, and 
combined this with the stellar radius derived in Section~\ref{sec:mcmc} to 
estimate the true stellar-rotation velocity, $v$ (Table~\ref{tab:sp}). 
For each star, the similarity of the values of $v$ and \vsini\ suggests the 
stellar spin axis is not significantly inclined relative to the sky plane.

\subsection{Photometric transits}
\label{sec:obs:phot}
The cadence of spectroscopic transit observations tends to be much lower 
than that of photometric observations and, at least for HARPS measurements of 
WASP systems, 
the spectroscopic exposure time tends to be similar to the durations of transit 
ingress and egress. 
As such, the accurate measurement of $\lambda$ requires the times of the four 
transit contact points during the spectroscopic transit to be accurately 
determined. 
This is best achieved using high-quality photometric transit observations. 
As M10 presented only a partial transit of WASP-22b and as the two 
transits of WASP-26b presented by S10 were obtained nine months prior to the 
RM measurement, we obtained additional transit photometry of the two systems.

\subsubsection{WASP-22b}
We used TRAPPIST\footnote{TRAnsiting Planets and PlanetesImals Small Telescope; 
http://arachnos.astro.ulg.ac.be/Sci/Trappist}, a 60-cm robotic telescope located 
at ESO La Silla \citep{2011EPJWC..1106002G} to observe two transits of WASP-22b. 
Both transits were observed through a special $I$+$z$ filter, in the 1$\times$2 
MHz read-out mode, and with 1$\times$1 binning, resulting in a typical combined 
readout and overhead time of 8 s and a readout noise of 13.5 $e^{-}$. 
With a pixel size of $15 \times 15$ $\mu$m, the pixel scale is 
0.65\arcsec\ pixel$^{-1}$ and the field of view is $22 \times 22$\arcmin.
The guiding system of TRAPPIST, based on absolute astrometry performed on 
each image, kept the stars at the same positions on the chip to within 
3\arcsec\ (5 pixels) over the course of the observations (Jehin et al., 2011).

The data from both transits were processed in the same way. After a standard 
pre-reduction (bias, dark and flat-field correction), the stellar fluxes were 
extracted from the images using the {\tt IRAF/DAOPHOT}\footnote{{\tt IRAF} is 
distributed by the National Optical Astronomy Observatory, which is operated by 
the Association of Universities for Research in Astronomy, Inc., under 
cooperative agreement with the National Science Foundation.} aperture photometry 
software \citep{1987PASP...99..191S}.  
We tested several sets of reduction parameters and chose the set that gave the 
most precise photometry for the stars of similar brightness to WASP-22. We 
carefully selected a set of reference stars and then performed differential 
photometry.

The first transit of WASP-22b that we observed occurred on 2010 Nov 18. We 
obtained little pre-transit data as the transit started soon after dusk. We also 
obtained little post-transit data as WASP-22 passed through the meridian soon 
after the transit ended, at which time a meridian flip would have been necessary 
as TRAPPIST uses a German equatorial mount. The post-flip light curve would 
thus have been of little value as, due to potential systematics, it would have 
needed to be detrended separately to the post-flip light curve. 
The second WASP-22b transit we observed occurred on 2010 Dec 2. We needed to 
perform a meridian flip shortly after beginning observations and shortly before
the transit began, so we divided the data into pre-flip and post-flip 
light curves. 

We also observed the 2010 Dec 2 transit of WASP-22b with EulerCam on the 
1.2-m Swiss telescope at ESO La Silla. 
Installed at the Swiss telescope in September 2010, EulerCam uses a 
4k $\times$ 4k, back-illuminated, deep-depletion $e2V$ silicon chip. 
Read-out of the entire chip takes $\sim$6 s in four-port mode and $\sim$25 s 
in one-port mode. In both modes the readout noise is 5 electrons.
The field of view of EulerCam is 15.6 $\times$ 15.6\arcmin, though the corners 
are shadowed by the filter wheel.
With a pixel size of 15 $\times$ 15 $\mu$m the pixel scale is 0.23\arcsec\ 
pixel$^{-1}$. 
EulerCam employs a very similar guiding system to that of TRAPPIST. 
Details of EulerCam will be provided in Lendl et al. (in preparation).

We observed WASP-22 for 4.8 hr through a Gunn $r$ filter with slight defocus 
and using one-port readout (as there were issues with four-port readout at the 
time).
Conditions were good, with a seeing of 0.5--1.0\arcsec. 
The target remained below airmass 1.8 as we observed from 35 min prior to the 
start of transit until 45 min after the transit ended. 
We bias-subtracted and flat-fielded the images before performing differential 
aperture photometry using a reference source created by co-adding the two most 
favourable reference stars. 

\begin{figure}[h!t]
\includegraphics[width=84mm]{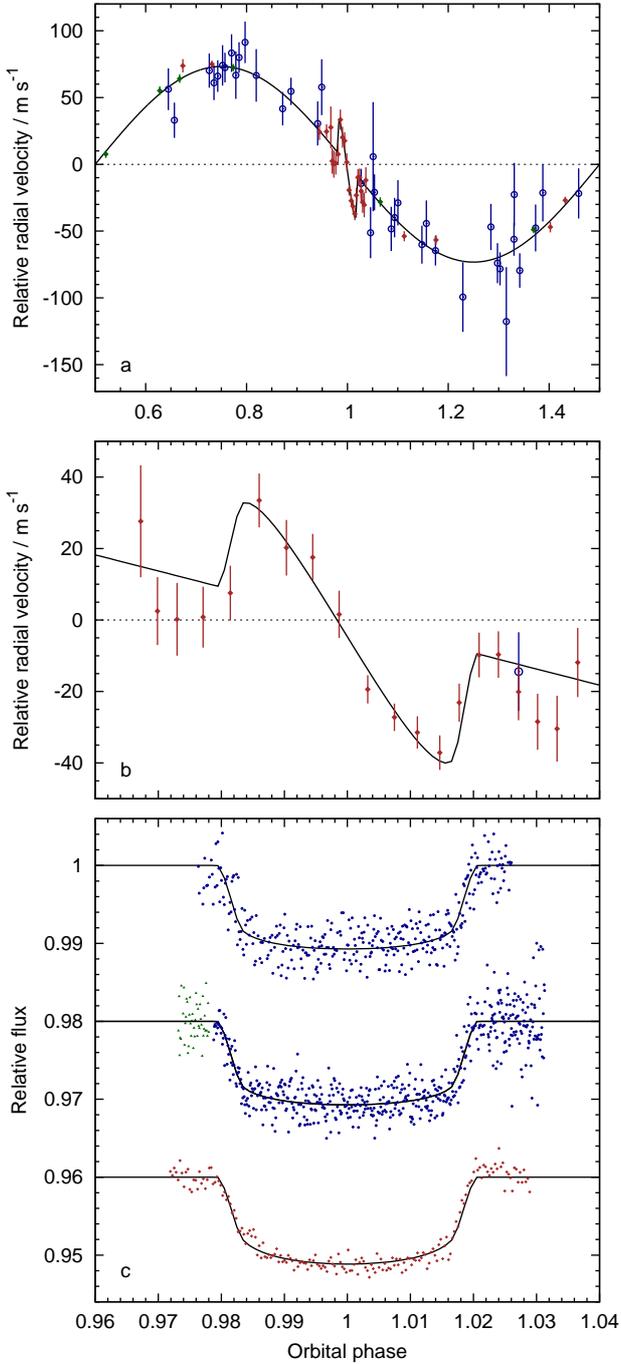}
\caption{{\bf (a)}: The CORALIE (blue circles) and HARPS (green diamonds) 
RVs of WASP-22 from M10, and the HARPS RVs from this paper (brown diamonds).
{\bf (b)}: An expansion of the region around the 
spectroscopic transit, as measured by HARPS on 2009 Nov 19. 
{\bf (c)}: The transit light curves, from top to bottom, obtained by TRAPPIST 
on 2010 Nov 18 and 2010 Dec 2 through an $I$+$z$ filter and by EulerCam on 2010 
Dec 2 through a Gunn $r$ filter. 
The data obtained by TRAPPIST on 2010 Dec 2 prior to the meridian flip are 
denoted by green triangles and blue dots denote the data obtained post-flip. 
In each panel, the data are phase-folded on the ephemeris given in 
Table~\ref{tab:mcmc} and the best-fitting models of Section~\ref{sec:mcmc} are 
superimposed. 
\label{fig:w22-rv-phot}} 
\end{figure}

The two transit light curves of WASP-22b from TRAPPIST and the one from Euler 
are shown in Figure~\ref{fig:w22-rv-phot}c. 
We also incorporated in our analysis of WASP-22 the photometry of M10: two 
seasons of WASP photometry covering the full orbit and a partial transit 
measured by Faulkes Telescope South through a Pan-STARRS $z$ filter.

\subsubsection{WASP-26b}
\begin{figure}[h!t]
\includegraphics[width=84mm]{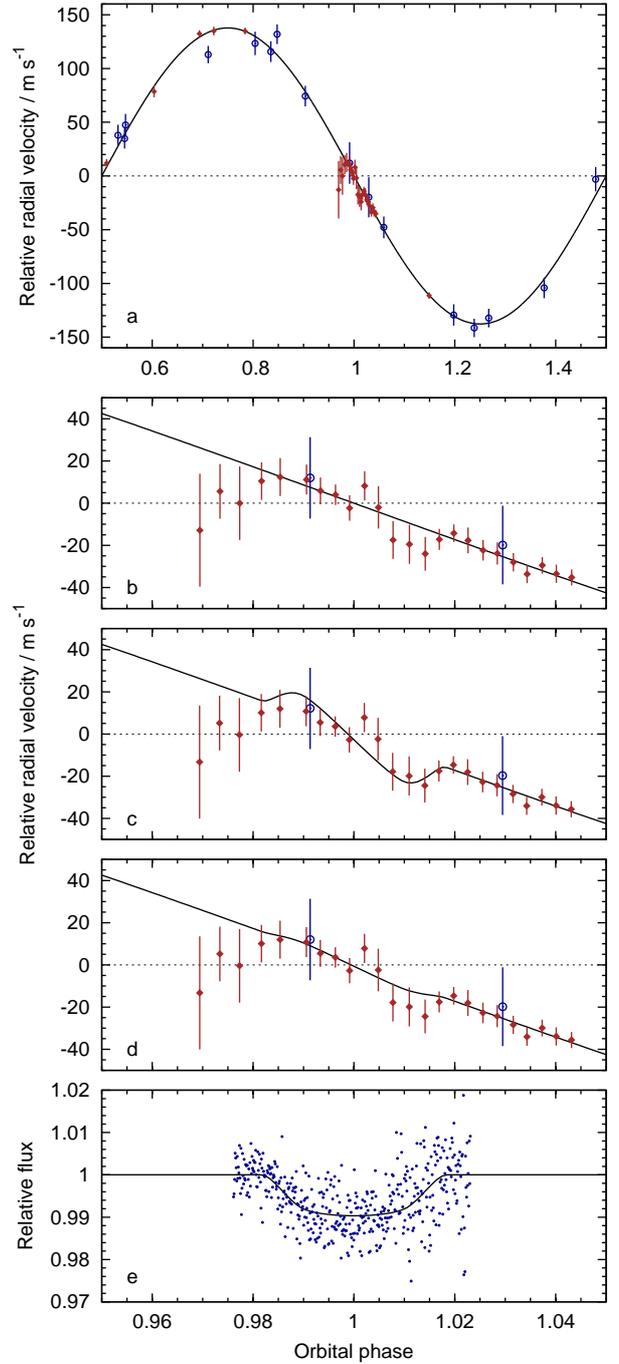}
\caption{{\bf (a)}: The CORALIE (blue circles) and HARPS 
(brown diamonds) RVs of WASP-26.
{\bf (b)}, {\bf (c)} and {\bf (d)}: Expansions of the region 
around the spectroscopic transit, as measured by HARPS on 2010 Sep 12, 
in which the RM effect is not modelled {\bf (b)}, 
the RM effect is modelled and a prior is imposed on \vsini\ {\bf (c)},
the RM effect is modelled without a prior on \vsini\ {\bf (d)}. 
{\bf (e)}: The Sloan $r'$-band transit light curve obtained at Oversky on 
2010 Oct 28/29.
\label{fig:w26-rv-phot}} 
\end{figure}

We observed a full transit of WASP-26b on 2010 Oct 28/29 through a Sloan $r'$ 
filter using the SBIG STL-1001E CCD camera installed at the prime focus of the 
0.35-m Celestron C14 Schmidt-Cassegrain telescope at the robotic Oversky 
Observatory, La Palma. 
The field of view is 19.9 $\times$ 19.9\arcmin\ and the exposure time was 15 s, 
with only a small degree of defocus employed so as to avoid blending with a 
star located 15\arcsec\ from the target. 
The data were processed using 
{\tt MUNIPACK}\footnote{\url{http://c-munipack.sourceforge.net}} and the transit 
light curve is shown in Figure~\ref{fig:w26-rv-phot}e. 

We also incorporated in our analysis of WASP-26 the photometry of S10: two 
seasons of WASP photometry covering the full orbit and two full transits 
measured through a Pan-STARRS $z$ filter, one by Faulkes Telescope North (FTN) 
and the other by Faulkes Telescope South (FTS). 
In-transit noise in the FTN light curve of S10 (see their Figure 4) appears to 
have affected its normalisation. Thus, the transit depth and, in turn, the 
stellar and plantetary radii may have been biased. As such we chose to 
detrend that light curve using only the post-transit data. 
S10 noted that WASP-26 is blended in the WASP data with the common-proper-motion 
star located 15\arcsec\ away that is $\sim$2.5 mag fainter. We made 
an approximate correction to the WASP photometry for this contamination so as to 
prevent dilution of the transit. 

\section{Combined model and analysis}
\label{sec:mcmc}

For each system, we determined the system parameters from a simultaneous fit to 
the data described in the previous section. 
The fit was performed using the current version of the Markov-Chain Monte Carlo 
(MCMC) code described by \citet{2007MNRAS.380.1230C} 
and \citet{2008MNRAS.385.1576P}.

The transit light curves were modelled using the formulation of 
\citet{2002ApJ...580L.171M} with the assumption that the planet is much smaller 
than the star. 
Limb-darkening was accounted for using a four-coefficient, nonlinear 
limb-darkening model, using coefficients appropriate to the passbands from the 
tabulations of \citet{2000A&A...363.1081C, 2004A&A...428.1001C}. 
The coefficients were interpolated once using the values of \logg\ and \feh\ 
from M10 and S10 (Table~\ref{tab:sp}), but were interpolated at each MCMC step 
using that step's value of \teff. The coefficient values corresponding to the 
best-fitting value of \teff\ are given in Table~\ref{tab:ld}.
The transit light curve is parameterised by the epoch of mid-transit 
$T_{\rm 0}$, the orbital period $P$, the planet-to-star area ratio 
(\rplanet/\rstar)$^2$, the duration of the transit from initial to 
final contact $T_{\rm 14}$, and the impact parameter $b = a \cos i/R_{\rm *}$ 
(the distance, in fractional stellar radii, of the transit chord from the 
star's centre in the case of a circular orbit), where $a$ is the semimajor axis 
and $i$ is the inclination of the orbital plane with respect to the sky plane. 
At each MCMC step, each light curve was decorrelated with a linear function of 
phase using singular value deconvolution. As noted in 
Section~\ref{sec:obs:phot}, we detrended the FTN transit of WASP-26b from S10 
using only the post-transit data, specifically those measurements with an 
orbital phase greater than 1.018.
From a fit to all data, this gives a slightly smaller (\rplanet/\rstar)$^2$ of  
$0.01022 \pm 0.00034$ as compared to the $0.01075 \pm 0.00033$ obtained when 
detrending using the entire light curve. This results in slightly smaller 
derived stellar and planetary radii 
($1.303 \pm 0.059$ \rsol\ c.f. $1.334 \pm 0.051$ \rsol\ 
and $1.281 \pm 0.075$ \rjup\ c.f. $1.346 \pm 0.067$ \rjup). 

The eccentric Keplerian radial-velocity orbit is parameterised by the stellar 
reflex velocity semi-amplitude $K_{\rm 1}$, the systemic velocity $\gamma$, 
an instrumental offset between the HARPS and CORALIE spectrographs 
$\Delta \gamma_{\rm HARPS}$, a linear drift in the systemic velocity 
$\dot\gamma$, and \secos\ and \sesin\ (Collier Cameron, in preparation),  
where $e$ is orbital eccentricity and $\omega$ is the argument of periastron. 
The RM effect was modelled using the formulation of \citet{2006ApJ...650..408G} 
and is parameterised by \svsicos\ and \svsisin.

The linear scale of the system depends on the orbital separation $a$ which, 
through Kepler's third law, depends on the stellar mass \mstar. 
At each step in the Markov chain, the latest values of \densstar, \teff\ and 
\feh\ are input in to the empirical mass calibration of 
\citet{2010A&A...516A..33E} to obtain \mstar.
The shapes of the transit light curves and the radial-velocity curve constrain 
\densstar\ \citep{2003ApJ...585.1038S}, which combines with \mstar\ to give 
\rstar.
\teff\ and \feh\ are proposal parameters constrained by Gaussian priors, by 
means of Bayesian penalties on $\chi^2$, with mean values and 
variances derived directly from the stellar spectra (Table~\ref{tab:sp}). 

As the planet-star area ratio is determined from the measured transit depth, 
\rplanet\ follows from \rstar. The planet mass \mplanet\ is calculated from the 
values of $K_{\rm 1}$ and \mstar, and planetary density \densplanet\ and surface 
gravity $\log g_{\rm pl}$ then follow. 
We also calculate the ingress and egress durations, $T_{\rm 12}$ and 
$T_{\rm 34}$, and the planetary equilibrium temperature \teql, assuming zero 
albedo and efficient redistribution of heat from the planet's presumed permanent 
day side to its night side.

At each step in the MCMC procedure, model transit light curves and radial 
velocity curves are computed from the proposal parameter values, which are 
perturbed from the previous values by a small, random amount. The $\chi^2$ 
statistic is used to judge the goodness of fit of these models to the data and a 
step is accepted if $\chi^2$ is lower than for the previous step. A step 
with higher $\chi^2$ is accepted with a probability 
proportional to $\exp(-\Delta \chi^2/2)$, 
which gives the procedure some robustness against local minima and leads to the 
thorough exploration of the parameter space around the best-fitting solution. 
To give proper weighting to each photometry dataset, the uncertainties were 
scaled at the start of the MCMC so as to obtain a photometric reduced-$\chi^2$ 
of unity. 
With a similar purpose, a jitter term of 4 \ms\ was added in quadrature to the 
formal errors of the HARPS RVs of WASP-22, as may be due to stellar 
activity (though there is no evidence of this here), correlated noise or the 
finite number of data-points.
This was determined from an initial MCMC from which the in-transit RVs were 
excluded. 
With spectroscopic reduced-$\chi^2$ values of less than unity, it was not 
necessary to add any jitter to the RVs of WASP-26.

To prevent a specific stellar activity level during a spectroscopic transit 
observation from biasing the fitting of the RM effect, RVs through a transit 
have previously been grouped separately to RVs spread over a long baseline 
\citep{2010A&A...524A..25T}. 
WASP-22 and WASP-26 both appear to be chromospherically inactive though (M10; 
S10), so we grouped RVs only by spectrograph so as to be able to determine the 
rate of drift in the systemic velocity whilst allowing for an instrumental 
offset. We did check that this choice did not affect the spin-orbit 
determination. 
The systemic velocity at the time of the spectroscopic transit is set by 
the RVs taken around that time, 
thus assisting in the accurate measurement of the projected spin-orbit angle. 
In the case of WASP-22, combining these HARPS RVs with those measured a year 
prior results in an improved determination of the linear drift in the systemic 
velocity noted in M10 and in tighter constraints on orbital eccentricity. 

For each system, the improvement in the fit to the RV data resulting from the 
use of an eccentric orbit model is small and is consistent with the 
underlying orbit being circular. 
We thus impose circular orbits, as is prudent to do for such short-period, 
Jupiter-mass planets in the absence of evidence to the contrary 
\citep[e.g.][]{2011arXiv1105.3179A}. This did not result in significant 
differences in the best-fitting values of the system parameters or their 
associated uncertainties. 
The 3-$\sigma$ upper limits on $e$ are 0.063 for WASP-22b and 0.050 for 
WASP-26b. 

\begin{figure}[h!]
\includegraphics[width=84mm]{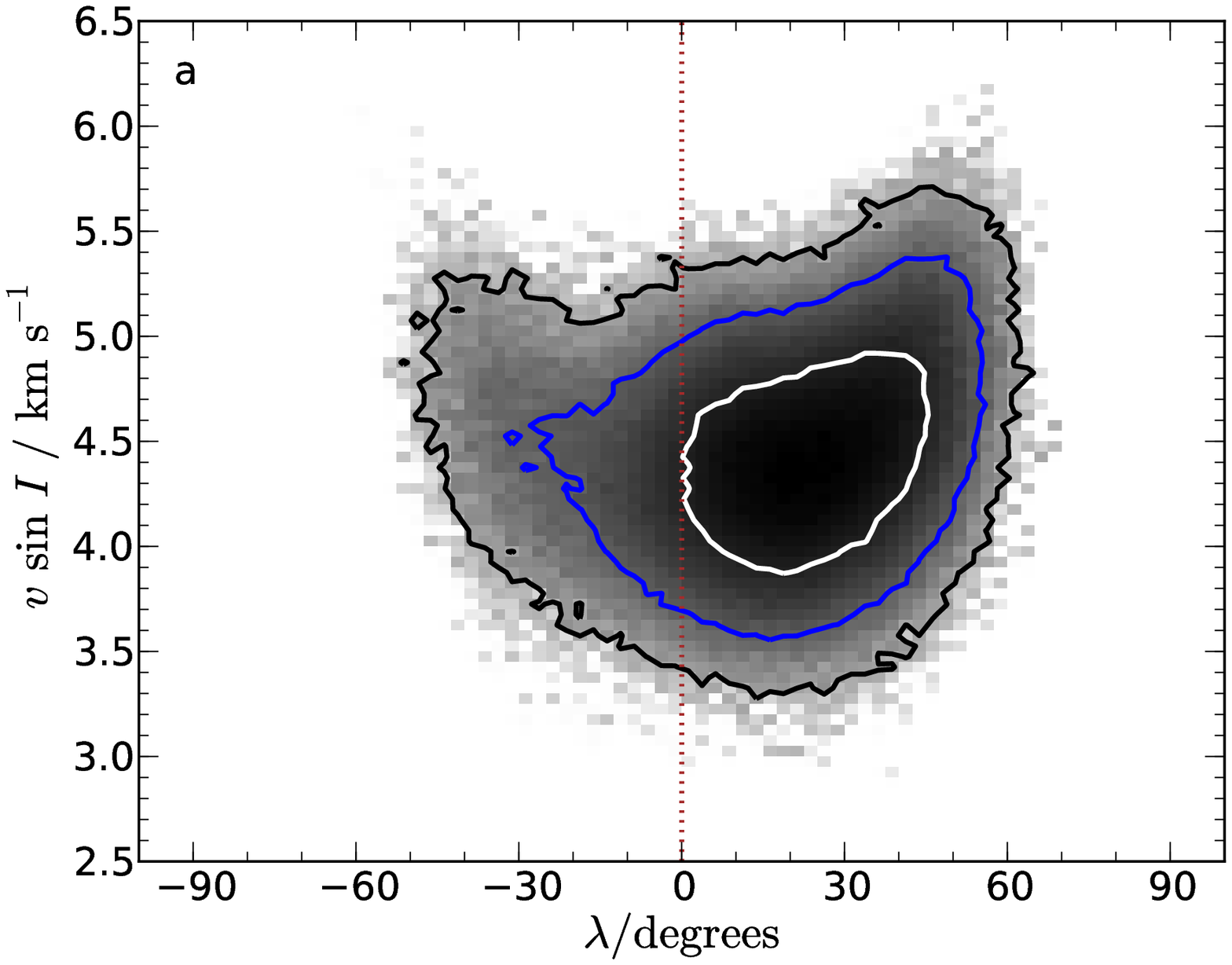} 
\includegraphics[width=84mm]{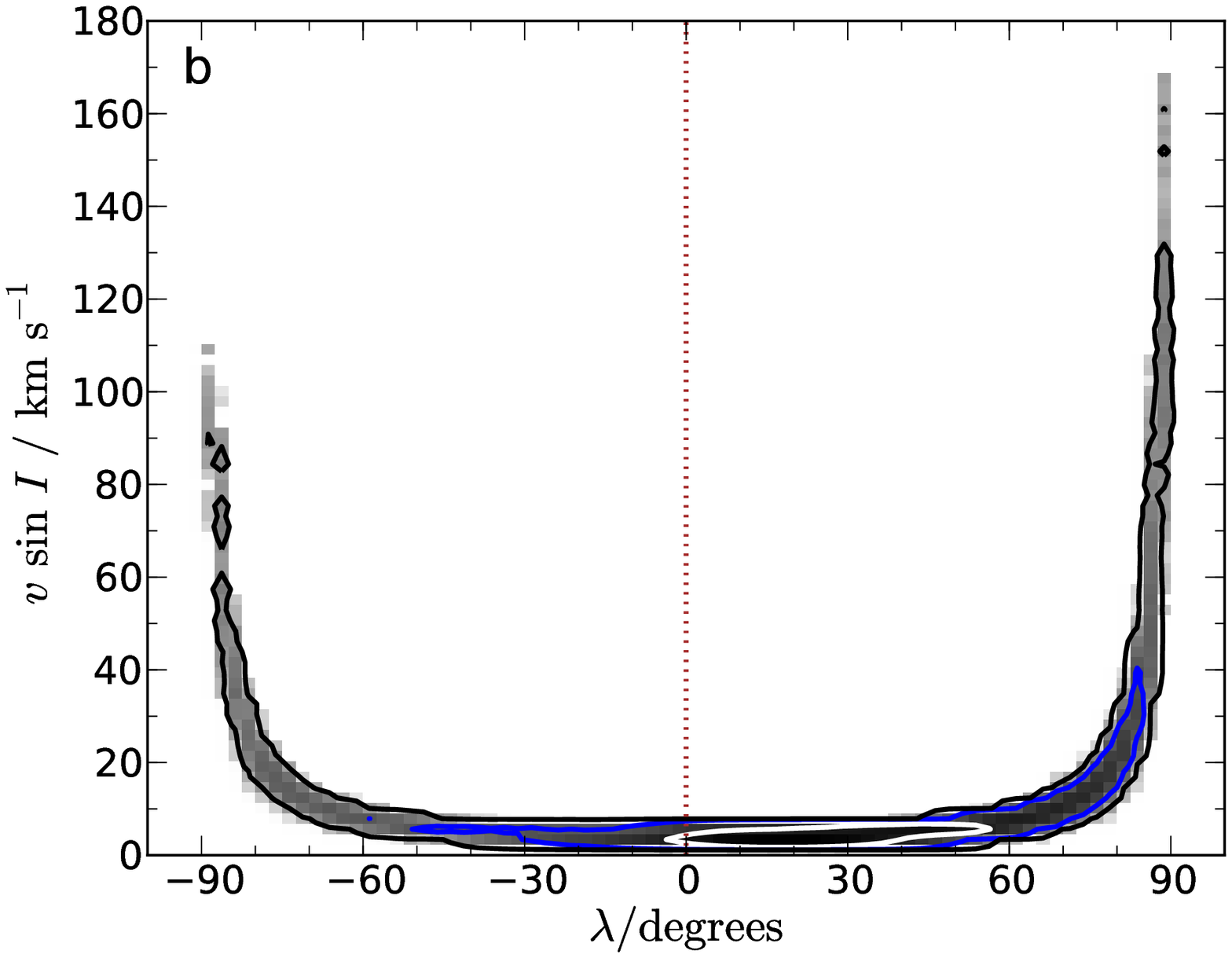} 
\caption{
The MCMC posterior distributions of $\lambda$ and \vsini\ for WASP-22 when 
{\bf (a)} a Gaussian prior of $4.5 \pm 0.4$ \kms is imposed on \vsini\ and 
{\bf (b)} no prior is imposed on \vsini. 
Note that, though the abscissa ranges of each graph are equal, the ordinate 
ranges are not. 
The white, blue and black contours are, respectively, the 1-, 2- and 3-$\sigma$ 
confidence limits. 
The shading of each bin is proportional to the logarithm of the number of MCMC 
steps within. 
Aligned spin-orbit axes would have $\lambda = 0^\circ$, which is marked by the 
dotted, brown line. 
\label{fig:lambda-vsini}} 
\end{figure}

We imposed a Gaussian prior on \vsini, by means of a Bayesian penalty on 
$\chi^2$, with mean and variance as determined from the HARPS spectra (see 
Section \ref{sec:obs:spec}). 
For WASP-22, this prior prevents \vsini\ from wandering to unrealistically large 
values of up to 168 \kms. 
This occurs due to a degeneracy between \vsini\ and $\lambda$ when $b$ is low: 
an RM effect with 
a similar amplitude and shape can result both when a planet transits 
near-parallel to the equator of a slowly rotating star and when a planet in a 
near-polar orbit transits a rapidly rotating star. 
The MCMC posterior distribution of \vsini\ and $\lambda$ shown in Figure 
\ref{fig:lambda-vsini}a results when a prior is imposed, and the distribution 
shown in Figure~\ref{fig:lambda-vsini}b results when no prior is imposed. 

We checked if the fitting of the RM effect, which involves the inclusion of the 
two parameters \vsini\ and $\lambda$, results in a significant improvement to 
the fit to the data for each system. 
For this we performed an F-test of the null hypothesis that we did not detect 
the RM effect, using a probability of $P(F) = 0.05$ as our significance 
threshold. 

For WASP-22, we find a highly significant improvement to the fit when modelling 
the RM effect with a prior on \vsini: $P(F) \sim 0$. 
So, we have made a significant detection of the RM effect of the WASP-22 system, 
with $\lambda = 22 \pm 16^\circ$, which is clearly visible in the star's RVs 
(Figure~\ref{fig:w22-rv-phot}b). 
Without the quadrature addition of the 4 \ms\ jitter term to the HARPS RVs, 
a value of $\lambda = 25 \pm 13^\circ$ is obtained. 

For WASP-26, we are unable to reject the null hypothesis as the data are fit 
equally well by both a model with no RM effect and by a model with the RM effect 
and a prior on \vsini, for which $\lambda = 6 \pm 6^\circ$ ($P(F) \sim 1$; see 
Figures~\ref{fig:w22-rv-phot}b and \ref{fig:w22-rv-phot}c). 
When modelling the RM effect with no prior on \vsini, for which 
$\lambda = 7 \pm 50^\circ$, we find the improvement in the fit is too slight 
to reject the null hypothesis ($P(F) = 0.50$; Figure~\ref{fig:w22-rv-phot}d). 
Also, the resulting best-fitting projected stellar rotation speed 
(\vsini\ = $0.7^{+1.1}_{-0.5}$ \kms) is discrepant with the spectral 
determination (\vsini\ = $3.9 \pm 0.4$ \kms). 
It can be seen in Figure~\ref{fig:w26-rv-phot}c that two in-transit RVs lie 
above the model. The seeing was poorer when those two measurements were 
obtained (1.1\arcsec) than during the rest of the RM sequence (0.5--0.8\arcsec). 
We explored excluding the four RVs (the two in-transit and two others 
outside of transit) with seeing greater than 1\arcsec, but found that a model 
with the RM effect and a prior on \vsini, for which $\lambda = 13 \pm 6^\circ$, 
was still not strongly favoured over a model with no RM effect: $P(F) = 0.09$. 
On the night of the RM measurement, three RVs were obtained prior to 
the re-focussing of the instrument. These data-points lie below the best-fitting 
model (Figure~\ref{fig:w26-rv-phot}). Having omitted these three RVs, a similar 
(slightly worse) fit to the RVs data resulted when fitting the RM effect with 
the prior on \vsini\ as compared to when not fitting the RM effect. 
As such, we did not make a significant detection of the RM effect for WASP-26 
and we adopt a Keplerian orbit with no RM effect as our model for that system. 
We note, though, that in Figure~\ref{fig:w26-rv-phot}b there appears to be 
some structure in the in-transit RVs.

We explored the degree to which the new transit photometry was assisting in 
the fitting of the RM effect by excluding it from MCMCs for the two systems.
For WASP-22b, the uncertainty in the transit mid-point at the epoch of the RM 
measurement was reduced from 3 to 1 minutes, and the uncertainty in the 
durations of the transit, ingress and egress fell from 6.5 to 1.4 minutes. 
For comparison, the ingress and egress durations are 20 minutes. 
Also reduced were the uncertainties in $b$ (from 0.16 to 0.12) and 
\rplanet/\rstar\ (from 0.024 to 0.013). 
These parameters, relevant to the fitting of the RM effect, and others were 
previously less certain due to the lack of available high-quality transit 
photometry. This led to the imposition of a main-sequence prior on the host star 
in M10, which was not imposed here in the MCMC omitting the new photometry. 
Without the new photometry $\lambda = 6^{+28}_{-12}$$^\circ$ is obtained. 
For WASP-26b, the uncertainty in the transit mid-point at the epoch of the RM 
measurement was reduced from 2 to 1 minutes, and the uncertainty in the 
durations of the transit, ingress and egress fell from 7.7 to 1.4 minutes. 
The ingress and egress durations are 34 minutes. 

By measuring the inclination of the stellar spin axis with respect to the 
sky plane $I$, we could determine the true, rather than the sky-projected, 
obliquity. 
A measurement of the stellar rotation period $P_{\rm rot}$, in combination with 
\rstar, would give the stellar rotation velocity $v$. 
By comparing this to the sky-projected rotation velocity \vsini, determined 
from the spectra, we could obtain $I$. 
$P_{\rm rot}$ can be measured if a star has an active photosphere that induces 
rotational modulation in its light curve \citep[e.g.][]{2011ApJ...730L..31H}. 
However, M10 found no evidence for modulation in the WASP-22 light curves, and 
S10 found none in the WASP-26 light curves.

\subsection{Results and discussion}

The results of the MCMC model fits are presented in Table~\ref{tab:mcmc}. 
The corresponding transit and orbit models are superimposed on the new transit 
light curves and on all RVs, respectively, in Figure~\ref{fig:w22-rv-phot} for 
WASP-22 and in Figure~\ref{fig:w26-rv-phot} for WASP-26. 

With $\lambda = 22 \pm 16^\circ$, WASP-22b is in a prograde orbit around its 
host star. The best-fitting model suggests that the planetary orbital axis is 
slightly misaligned with the stellar spin axis, but the current uncertainty 
allows for quite a range in $\lambda$ (Figure~\ref{fig:lambda-vsini}a).
The similarity of our estimates of \vsini\ and $v$ (Table~\ref{tab:sp}), which  
agrees with the prediction of \citet{2010ApJ...719..602S} that $v = 3.2 \pm 1.0$ 
\kms, suggests that the spin axis of WASP-22 is not significantly inclined 
relative to the sky plane (i.e. $\Psi \sim \lambda$). 
With \teff\ = $6000 \pm 100$ K, WASP-22 is consistent with the observation of 
\citet{2010ApJ...718L.145W} that the orbits of planets around stars cooler than 
$\sim$6250 K tend to be aligned. 
Conversely, the orbits of planets around hotter stars tend to be misaligned, 
as noted by both \citet{2010ApJ...718L.145W} and \citet{2010ApJ...719..602S}. 
A few possible expections to this trend are now known, including HAT-P-9 
with \teff\ = 6350 K and $\lambda = -16 \pm 8^\circ$ 
\citep{2011A&A...533A.113M}. Though it may be that such systems were never very 
misaligned. 
\citet{2010ApJ...718L.145W} suggested the observed trend either indicates that 
planetary formation and migration mechanisms depend on stellar mass, or that the 
larger convective envelopes of cooler stars are able to more effectively realign 
with orbits via tidal dissipation, though this does depend on speculative, 
long-lived core-envelope decoupling.

\begin{figure}[h!]
\includegraphics[width=84mm]{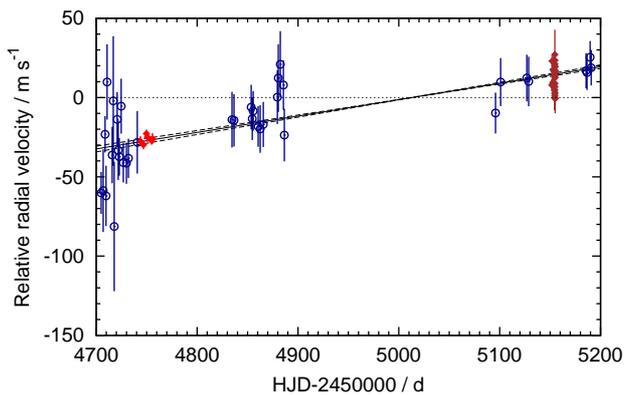}
\caption{Radial velocities of WASP-22 showing a linear drift in the systemic 
velocity of $\dot\gamma = 37.8 \pm 2.2$ \ms\ yr$^{-1}$. 
The colour and symbol key is the same as in Figure~\ref{fig:w22-rv-phot}.
The Keplerian orbit of Table~\ref{tab:mcmc} was subtracted. 
The value of $\dot\gamma$ from Table~\ref{tab:mcmc} is plotted as a solid line, 
relative to JD = 2\,455\,013, which is the centre-of-mass of the RV data, 
weighted by the square of the measurement precision. 
The drift's 1-$\sigma$ error bars are plotted as dashed lines. 
\label{fig:w22-drift}} 
\end{figure}

M10 detected a linear trend of $\dot\gamma = 40 \pm 5$ \ms\ in the RVs of 
WASP-22, which span a duration of 16 months. 
The new HARPS RVs presented herein make possible a more precise determination 
of the trend of $\dot\gamma = 37.8 \pm 2.2$ \ms\ (Figure~\ref{fig:w22-drift}). 
We can obtain an order of magnitude estimate of the minimum mass of the third 
body, $M_3 \sin i_3$, scaled by its orbital distance, $a_3$, 
from the primary. 
Following \citet{2010ApJ...718..575W}, by assuming the third body to be in a 
near-circular orbit and to have a mass much smaller than that of the primary, we 
may set $\dot\gamma \sim G M_3 \sin i_3 / a_3^2$ to obtain: 

\begin{align}
\left(\frac{M_3 \sin i_3}{\mjup}\right) 
\left(\frac{a_3}{5{\rm AU}}\right)^{-2} 
\sim  5.3 \pm 0.3.
\end{align}

The RM effect of WASP-26b, predicted to have a semi-amplitude of $\sim$15 \ms, 
was not detected with significance in our data, for which the median in-transit 
uncertainty was 7 \ms. This was due to observation noise and the small amplitude 
of the signal. 
For comparison, we measured the semi-amplitude of the RM effect of WASP-22b to 
be $\sim$30 \ms, which agrees well with our prediction of 27 \ms. The median 
in-transit RV uncertainty was 8 \ms. 
The predicted amplitude of the RM effect of WASP-26b is smaller than that of 
WASP-22b due to the higher impact parameter and, to a lesser extent, the slower 
stellar rotation velocity. 
The similarity of our estimates of \vsini\ and $v$ (Table~\ref{tab:sp}), which 
agrees with the prediction of \citet{2010ApJ...719..602S} that $v = 3.1 \pm 0.3$ 
\kms, suggests that the spin axis of WASP-26 is not significantly inclined 
relative to the sky plane. 
Thus it is unlikely that the non-detection of the RM effect is due to the 
orientation of WASP-26 being close to pole-on. 
Though it is possible that the planet is in a polar orbit, it can not be 
transiting over the stellar spin axis as the impact parameter is non-zero. 
Hence, there will be an observable RM signal and thus further measurements are 
warranted. 

The available RVs of WASP-26 are not a very sensitive probe of a long-term drift 
as the CORALIE RVs span only 95 days and the HARPS RVs span only 5 days, with a 
gap of 1 year between the two datasets. 
Any drift could therefore have been taken up within the fitted instrumental 
offset between the two spectrographs. 
However, the similarity in the offsets for the two systems (Table~\ref{tab:mcmc}) 
suggests that any drift in the systemic velocity of WASP-26 is likely to be 
smaller than that of WASP-22.

\begin{table*}
\centering
\caption{Limb-darkening coefficients} 
\label{tab:ld} 
\begin{tabular}{lccccccc}
\hline
System		& Instrument	& Observation band		& Claret band	& $a_1$		& $a_2$		& $a_3$		& $a_4$	\\
\hline
WASP-22		& WASP / Euler	& Broad (400--700 nm) / Gunn $r$& Cousins $R$	& 0.578		& $-$0.049	& 0.494		& $-$0.294\\
WASP-22		& TRAPPIST	& Cousins $I$+Sloan $z'$	& Sloan $z'$	& 0.652		& $-$0.348	& 0.638		& $-$0.326\\
WASP-26		& WASP		& Broad (400--700 nm)		& Cousins $R$	& 0.573		& $-$0.027	& 0.464		& $-$0.281\\
WASP-26		& FTN / FTS	& Pan-STARRS $z$		& Sloan $z'$	& 0.651		& $-$0.337	& 0.624		& $-$0.321\\
WASP-26		& Oversky	& Sloan $r'$			& Sloan $r'$	& 0.551		& ~~\,0.048		& 0.415		& $-$0.266\\
\hline
\end{tabular}
\end{table*}

\begin{table*} 
\centering
\caption{System parameters from RV and transit data from our adopted, circular 
solutions} 
\label{tab:mcmc}
\begin{tabular}{lcccc}
\hline
Parameter & Symbol & WASP-22 & WASP-26 & Unit\\ 
\hline 
\\
Orbital period			& $P$	 			& $3.5327313 \pm 0.0000058$		& $2.7566004 \pm 0.0000067$ 	& d\\
Epoch of mid-transit (HJD,UTC)	& $T_{\rm 0}$	 		& $2455497.39967 \pm 0.00025$		& $2455228.38842 \pm 0.00058$	& d\\
Transit duration		& $T_{\rm 14}$	 		& 0.14501$^{+ 0.00114}_{- 0.00087}$	& $0.0993 \pm 0.0018$		& d\\
Transit ingress (= egress) duration		& $T_{\rm 12}=T_{\rm 34}$ 	& 0.01369$^{+ 0.00124}_{- 0.00074}$ 	& $0.0239 \pm 0.0026$		& d\\
Planet-star area ratio		& $(R_{\rm P}$/R$_{*})^{2}$	& $0.00954 \pm 0.00018$			& $0.01022 \pm 0.00034$		&  \\
Impact parameter		& $b$ 				& $0.25 \pm 0.12$ 			& $0.812 \pm 0.020$ 		& \\
Orbital inclination \medskip	& $i$  				& $88.26 \pm 0.91$ 			& $82.91 \pm 0.46$ 		& $^\circ$\\

Stellar reflex velocity semi-amplitude		& $K_{\rm 1}$	& $73.2 \pm 1.7$			& $137.7 \pm 1.5$		& \ms\\
Semi-major axis					& $a$		& $0.04698 \pm 0.00037$			& $0.03985 \pm 0.00033$		& AU\\
Offset between HARPS and CORALIE& $\Delta \gamma_{\rm HARPS}$	& $17.60 \pm 0.80$			& $20.62 \pm 0.14$		& \ms\\
Systemic velocity at time $T_{\rm 0}$		& $\gamma$			& $-7\,187.3 \pm 3.5$ 			& $8\,459.297 \pm 0.073$	& \ms\\
Linear drift in systemic velocity & $\dot\gamma$		& $37.8 \pm 2.2$			& ---				& \ms yr$^{-1}$\\
Orbital eccentricity 		& $e$ 				& 0 (adopted) 				& 0 (adopted)			& \\
		\medskip	&				& $<0.063$ (3 $\sigma$)			& $<0.048$ (3 $\sigma$)		& \\
								
Sky-projected spin-orbit angle 			& $\lambda$	& $22 \pm 16$				& ---				& $^\circ$\\
Sky-projected stellar rotation velocity 	& \vsini\	& $4.42 \pm 0.34$			& ---				& \kms\\
						& \vsini\ $\cos \lambda$ & $3.98 \pm 0.43$		& ---				& \kms\\
\medskip					& \vsini\ $\sin \lambda$ & $1.63 \pm 1.13$		& ---				& \kms\\

Stellar mass					& \mstar	& $1.109 \pm 0.026$			& $1.111 \pm 0.028$		& \msol\\
Stellar radius					& \rstar	& 1.219$^{+ 0.052}_{- 0.033}$		& $1.303 \pm 0.059$		& \rsol\\
Stellar surface gravity				& \logg		& 4.310$^{+ 0.020}_{- 0.032}$		& $4.253 \pm 0.034$		& (cgs)\\
Stellar density					& \densstar	& 0.612$^{+ 0.047}_{- 0.067}$		& $0.502 \pm 0.062$		& \denssol\\
Stellar effective temperature			& \teff		& $5958 \pm 98$				& $5939 \pm 100$		& K\\
Stellar metallicity \medskip 			& \feh		& $0.050 \pm 0.080$ 			& $-0.020 \pm 0.091$		& \\

Planetary mass 					& \mplanet	& $0.588 \pm 0.017$			& $1.028 \pm 0.021$		& \mjup\\
Planetary radius				& \rplanet	& 1.158$^{+ 0.061}_{- 0.038}$		& $1.281 \pm 0.075$		& \rjup\\
Planetary surface gravity			& \loggplanet	& 3.001$^{+ 0.028}_{- 0.041}$		& $3.156 \pm 0.048$		& (cgs)\\
Planetary density				& \densplanet	& 0.378$^{+ 0.038}_{- 0.051}$		& $0.488 \pm 0.082$		& \densjup\\
Planetary equilibrium temperature		& \teql		& $1466 \pm 34$				& $1637 \pm 45$			& K\\
\\ 
\hline 
\end{tabular} 
\end{table*} 

\begin{acknowledgements} 
TRAPPIST is a project funded by the Belgian Fund for Scientific Research (Fond 
National de la Recherche Scientifique, FNRS) under the grant FRFC 2.5.594.09.F, 
with the participation of the Swiss National Science Fundation (SNF).
M. Gillon and E. Jehin are FNRS Research Associates.
\end{acknowledgements}


\end{document}